\documentclass[12pt]{ecoc01}

\usepackage{amsmath,amssymb}

\usepackage{mathptmx,graphicx}
\usepackage{caption,subcaption,colordvi,psfrag}
\usepackage{epstopdf}
\usepackage{calc,pstricks, pgf, xcolor}
\usepackage{cite}
\usepackage{epsfig}
\usepackage{setspace}
\newcommand{\abs}[1]{\left\vert#1\right\vert}

\newcommand{\SNR}{\text{SNR}}
\newcommand{\be}{\begin{equation}}
\newcommand{\ee}{\end{equation}}
\newcommand{\bea}{\begin{eqnarray}}
\newcommand{\eea}{\end{eqnarray}}
\begin{document}

\title{New Bounds on the Capacity of Fiber-Optics Communications}

\markboth{Ronen Dar, Mark Shtaif, Meir Feder}{New Bounds on the Capacity of Fiber-Optics Communications}

\author{Ronen Dar$^{(1)}$, Mark Shtaif$^{(2)}$, Meir Feder$^{(3)}$}

\address{School of Electrical Engineering, Tel Aviv University, Tel Aviv 69978, Israel\\
$^{(1)}${ronendar@post.tau.ac.il}, $^{(2)}${shtaif@tauex.tau.ac.il}, $^{(3)}${meir@eng.tau.ac.il}}

\begin{abstract}
By taking advantage of the temporal correlations of the nonlinear phase noise in WDM systems we show that the capacity of a nonlinear fiber link is notably higher than what is currently assumed. This advantage is translated into the doubling of the link distance for a fixed transmission rate.
\vspace{1pc}
\end{abstract}

\twocolumn

\maketitle

\section{Introduction}
Estimation of the fiber-optic channel capacity has come to be one of the most challenging and important problems in the field of optical communications \cite{mitraNonlinearCapacity,TuritsynInformationCapacityOfOpticalFiber,CapacityLimitsofOpticalFiberNetworks,ellisApproachingTheNonLinearShannonLimit,BoscoAnalyticalResultsOnChannelCapacity,PseudolinearCapacityMecozzi,agrellCapacity}. Recently, its importance has grown to be even bigger as the latest capacity estimates are being rapidly approached by the rates of commercial communications systems \cite{WinzerFoschiniOpticalMIMO}. The difficulty in estimating the capacity of the fiber-optic channel is mostly due to the effect of fiber nonlinearity which generates complicated distortions of the transmitted optical waveforms. Perhaps the most comprehensive and familiar attempt of estimating the fiber-channel capacity to date is due to Essiambre et al. \cite{CapacityLimitsofOpticalFiberNetworks}, where it was argued that, under plausible assumptions on network architecture, nonlinear interference between different wavelength-division multiplexed (WDM) channels must be treated as noise, which was then identified as the predominant nonlinear factor in limiting the capacity of the fiber-optic channel. This point of view has been adopted by the subsequent studies \cite{BoscoAnalyticalResultsOnChannelCapacity,PseudolinearCapacityMecozzi}, and we also adopt it in the study presented herein.

A common feature of capacity estimates published so far is that they treat the nonlinear noise as additive, white and independent of the data transmitted on the channel of interest. In reality, in the presence of chromatic dispersion, different WDM channels propagate at different velocities so that every symbol in the channel of interest interacts with multiple symbols of every interfering channel. Consequently, adjacent symbols in the channel of interest are disturbed by essentially the same collection of interfering pulses and therefore they are affected by nonlinearity in a highly correlated manner. In addition, as has been recently demonstrated by Mecozzi et al. \cite{PseudolinearCapacityMecozzi}, the most pronounced manifestation of nonlinearity is in the form of phase noise due to cross-phase-modulation (XPM).

We demonstrate in what follows that by taking advantage of these properties, it is possible to communicate at a higher rate than predicted in \cite{CapacityLimitsofOpticalFiberNetworks}, or equivalently (almost) double the distance achievable at a given rate of communications.

\section{Channel capacity with correlated phase noise}
We express the received signal samples after coherent detection and matched filtering as\newline
\be y_j=x_j\exp(i\theta_j)+n^{\mathrm{NL}}_j+n_j, \label{hjf98329}\ee
where the term $n^{\mathrm{NL}}_j$ accounts for all nonlinear noise contributions that do not manifest themselves as phase noise.  As in \cite{CapacityLimitsofOpticalFiberNetworks,ellisApproachingTheNonLinearShannonLimit,BoscoAnalyticalResultsOnChannelCapacity,PseudolinearCapacityMecozzi} we assume that the samples $n^{\mathrm{NL}}_j$ are statistically independent complex Gaussian variables with variance $\sigma_{\mathrm{NL}}^2$. A similar assumption holds for the amplified spontaneous emission (ASE) samples $n_j$, whose variance is denoted by $\sigma_{\mathrm{ASE}}^2$. All three noise contributions $\theta_j$, $n^{\mathrm{NL}}_j$, and $n_j$ are assumed to be statistically independent of each other. All of the above assumptions, regarding the whiteness of $n^{\mathrm{NL}}_j$ and $n_j$, the statistical independence of all noise contributions and their Gaussianity constitute a {\bf\textit{worst case}} in terms of the resultant capacity \cite{CoverElementsOfInformationTheory} and hence they are in accord with our goal of deriving a capacity lower bound. Finally, consistently with what is suggested by the analysis in \cite{PseudolinearCapacityMecozzi}, we will also assume that $\theta_j$ is a Gaussian distributed variable and its variance will be denoted by $\sigma_\theta^2$. Since we show that the phase noise essentially can be canceled entirely, this assumption has almost no impact the results presented here. Now, for arriving at an analytical lower bound for the capacity, we assume that the nonlinear phase-noise $\theta_j$ is blockwise constant. In other words, it is assumed that the noise $\theta_j$ does not change at all within a block of $N$ symbols and then in the subsequent block it changes in a statistically independent manner. The assumption of statistical independence of $\theta_j$ in adjacent blocks is again a worst-case scenario which is in accord with our interest in a lower bound.

\begin{figure}[h]
\center
\epsfig{file=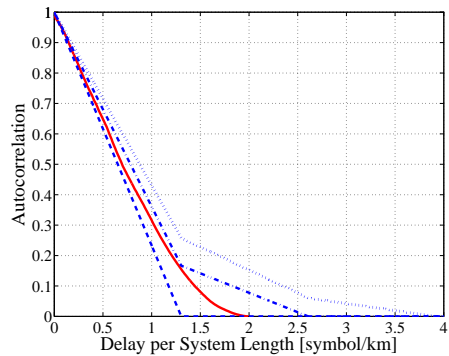,height=0.3\textwidth,width=0.4\textwidth,clip=}
\caption{The autocorrelation function of the phase noise $\theta_j$. The delay is normalized to the length of the link and expressed in symbols per km. The dashed, dash-dotted and dotted curves show the analytical result (based on \cite{PseudolinearCapacityMecozzi}) for cases in which the channel of interest is surrounded by 1,2, and 3 WDM channels on each side, respectively. The red curve was extracted numerically  from our simulations for the case of 5 WDM channels (2 neighbors on each side).}
\label{fig2eqw23}
\end{figure}

The assumption of constant phase noise within a block of symbols is well justified in view of the very long temporal correlation of the phase. In order to demonstrate this we extend the analysis of \cite{PseudolinearCapacityMecozzi} to obtain the autocorrelation function (ACF) of the phase noise, assuming an equal input average power in all channels,
%
\bea
R(m)&=&\frac{\mathbb E [\theta_j\theta_{j+m}]}{\sigma_\theta^2}\nonumber\\
&=&\sum_{s=1}^{N_{ch}}\frac{c_s}{\sum_{s'=1}^{N_{ch}}c_{s'}}\big[1-\frac{\abs{m}}{L}L_{wo_s}\big]^+, \label{fewfew}
\eea
where $\mathbb E$ denotes statistical averaging and $[a]^+$ denotes for $\max\{0,a\}$. The term $N_{ch}$ is the number of interfering channels, $L$ is the transmission distance, $L_{wo_s}=T/\abs{\beta_2 \omega_s}$ is the walk-off distance of the $s$-th interfering channel, $T$ is the time delay between pulses, $\omega_{s}$ is the frequency spacing between the $s$-th channel and the channel of interest, $c_s=(k_s-1)/\abs{\omega_s}$ where $k_s={\mathbb{E}(\abs{x_{s}}^4)}/{\mathbb{E}(\abs{x_{s}}^2)^2}$ ($x_s$ represents the information carrying symbol transmitted through the $s$-th channel). Note that when $k_s=1$ the information carrying symbols on the $s$-th channel are phase modulated only, and as is shown in \cite{PseudolinearCapacityMecozzi}, induce zero phase fluctuations.
The theoretical ACF is plotted in Fig. \ref{fig2eqw23} for the case of 1, 2 and 3 interfering channels from each side of the channel of interest with Gaussian modulation (i.e. $k_s=2$) in all channels. The rest of the parameters are given in section 3. Notice for example that in a 1000 km link, the 3dB width of the ACF is roughly equal to 700 symbols, whereas for 100 symbols the ACF remains very close to unity, implying that $\theta_j$ remains practically constant on this time scale.

The theoretical variance of the phase noise, as was presented in \cite{PseudolinearCapacityMecozzi}, is given here for completeness,
\be \sigma_\theta^2= 4\gamma^2 T L \sum_{s=1}^{N_{ch}}\frac{k_s-1}{\abs{\beta_2\omega_s}}P_s^2~, \label{eqcsdcf834} \ee
where $\gamma$ is the nonlinearity coefficient and $P_s$ is the average signal power lunched through the $s$-th interfering channel.

The capacity of the block-wise independent phase noise channel \eqref{hjf98329} is given by
\be \mathrm{C}=\frac{1}{N} \sup_{p(\underline{x})} ~{{\textit {I}}} (\underline{x};\underline{y}), \ee
where $\underline{x}$ and $\underline{y}$ are column vectors representing a block of $N$ channel inputs and outputs, respectively, in which the phase noise is constant. The supremum is over all input distributions satisfying the power constraint $\mathbb{E}[\| \underline{x} \|^2]=NP$. A lower bound is obtained by considering statistically independent circularly-symmetric complex Gaussian inputs. In this case, $\underline{y}$ is circularly symmetric complex Gaussian vector with differential entropy $h(\underline{y})=N\log_2(\pi e(P+\sigma_{\mathrm{eff}}^2))$, where $\sigma_{\mathrm{eff}}^2=\sigma_{\mathrm{ASE}}^2+\sigma_{\mathrm{NL}}^2$. Since normal distribution maximizes the differential entropy of a vector of zero-mean random variables with a given covariance matrix, $h(\underline{y}|\underline{x})$ satisfies
\begin{eqnarray}
h(\underline{y}|\underline{x})&=&\mathbb{E}_{\underline{x}}\big(h(\underline{y}|\underline{x})\big) \\
&\leq & \frac{1}{2}\mathbb{E}_{\underline{x}} \big(\log_2 \det( 2\pi e ~Q_{\underline{\hat{y}}| \underline{x}})\big) ~, \label{eq1010}
\end{eqnarray}
where $\underline {\hat{y}}=\left[\begin{smallmatrix}\mathfrak{Re}(\underline {y}) \\\mathfrak{Im}(\underline {y})\end{smallmatrix}\right]$ and $Q_{\underline {\hat{y}}| \underline x}$ is the covariance matrix of $\underline {\hat{y}}$ given $\underline x$. By applying some algebraic manipulations the determinant of $Q_{\underline{\hat{y}}| \underline{x}}$ can be shown to satisfy
\begin{align}
\det(Q_{\underline{\hat{y}}| \underline{x}})&=(\frac{\sigma^2_{\mathrm{eff}}}{2})^{2N} (1+2\frac{\|\underline{x}\|^2}{\sigma^2_{\mathrm{eff}}}\sigma^2_c)(1+2\frac{\|\underline{x}\|^2}{\sigma^2_{\mathrm{eff}}}\sigma^2_s)~, \label{eq11030}
\end{align}
where the terms $\sigma^2_c=0.5(1-e^{-\sigma_\theta^2})^2$ and $\sigma^2_s=0.5(1-e^{-2\sigma_\theta^2})$ are the variances of $\cos(\theta)$ and $\sin(\theta)$, respectively. Plugging \eqref{eq11030} into \eqref{eq1010}, the following capacity lower bound is obtained
\bea
\mathrm{C}&\geq&\log_2\left(1+\frac{P}{\sigma_{\mathrm{eff}}^2}\right)\nonumber\\
&-&\frac{1}{2N}\mathbb{E}_\upsilon \left\{\log_2 \left(1+\upsilon \sigma^2_c\frac{P}{\sigma_{\mathrm{eff}}^2}\right)\right\} \nonumber\\
&-&\frac{1}{2N}\mathbb{E}_\upsilon \left\{\log_2 \left(1+\upsilon \sigma^2_s \frac{P}{\sigma_{\mathrm{eff}}^2}\right)\right\}, \label{eq13900a}
\eea
where the symbol $\mathbb{E}_\upsilon$ stands for ensemble averaging with respect to a standard Chi-square distributed variable $\upsilon$ with $2N$ degrees of freedom (DoF). Notice that the first line on the right-hand-side of Eq. (\ref{eq13900a}) is identical to the result of \cite{CapacityLimitsofOpticalFiberNetworks}, except that in our case $\sigma^2_{\mathrm{NL}}$ accounts only for the part of the nonlinear noise that does not manifest itself as phase-noise and hence it is smaller than the corresponding term appearing in \cite{CapacityLimitsofOpticalFiberNetworks}. The effect of phase-noise on the capacity is captured in our case by the bottom two lines on the right-hand-side of (\ref{eq13900a}). This capacity loss, which may be viewed as a rate reduction needed for estimating the phase noise, vanishes when the phase exhibits very long term correlations.

The capacity of block-wise independent phase noise channel, although unknown in general, is approximately $\left(1-\tfrac{1}{2N}\right)\log_2\left(\SNR\right)$ for high SNR \cite{LapidothCapacityBoundsViaDuality}; that is, out of $2N$ DoF available in a transmission of $N$ complex symbols, one DoF is lost due to the phase noise. According to our capacity bound \eqref{eq13900a}, two DoF are lost when $P$ goes to infinity (letting $\sigma^2_{\mathrm{eff}}$ be independent of $P$, just for a theoretical discussion on the tightness of our bound). We have derived tighter bounds for the high SNR regime, however omitted them from this paper as in the intermediate regime of SNR values, where the maximal optical-fiber capacity is obtained, the bound given by \eqref{eq13900a} attains the best result. 

\section{Simulation and results}


The goal of the simulations we conducted was to validate the character of the nonlinear phase-noise and to extract the parameters $\sigma^2_\theta$ and $\sigma_{\mathrm{eff}}^2$.
The simulations were performed using the parameters of a standard single mode fiber; dispersion $D= 17$ ps/nm/km, attenuation of 0.2 dB/km, nonlinear coefficient $\gamma=1.27$ W$^{-1}$km$^{-1}$ and signal wavelength $\lambda_0=1.55~\mu $m. Perfectly distributed and quantum limited (i.e. fully inverted) amplification with spontaneous emission factor $n_{sp}=1$ was assumed. Sinc-shaped pulses with a perfectly square 100 GHz wide spectrum were used for transmission and the spacing between adjacent WDM channels was 102 GHz (i.e leaving a 2 GHz guard band). The number of simulated WDM channels was 5, with the central channel being the channel of interest. All of the above assumptions are identical to those made by Essiambre et al. in \cite{CapacityLimitsofOpticalFiberNetworks}. The number of simulated symbols in each run was 8192 for 500 km system and 16384 for 1000 km and 2000 km systems. Up to 500 runs (each with independent and random data symbols) were performed with each set of system parameters, so as to accumulate sufficient statistics.
We assumed a circularly symmetric complex Gaussian distribution of points in the transmitted constellation. This constellation was used to derive our capacity lower bound \eqref{eq13900a}.
At the receiver, the central channel is filtered out with a perfectly square filter (which is also the matched filter with sinc pulses) and back-propagated. Then, the signal is appropriately sampled and analyzed. As in \cite{CapacityLimitsofOpticalFiberNetworks}, all simulations have been performed with the scalar nonlinear Schr\"{o}dinger equation and correspond to a single polarization.

\begin{figure}[h]
\center
\epsfig{file=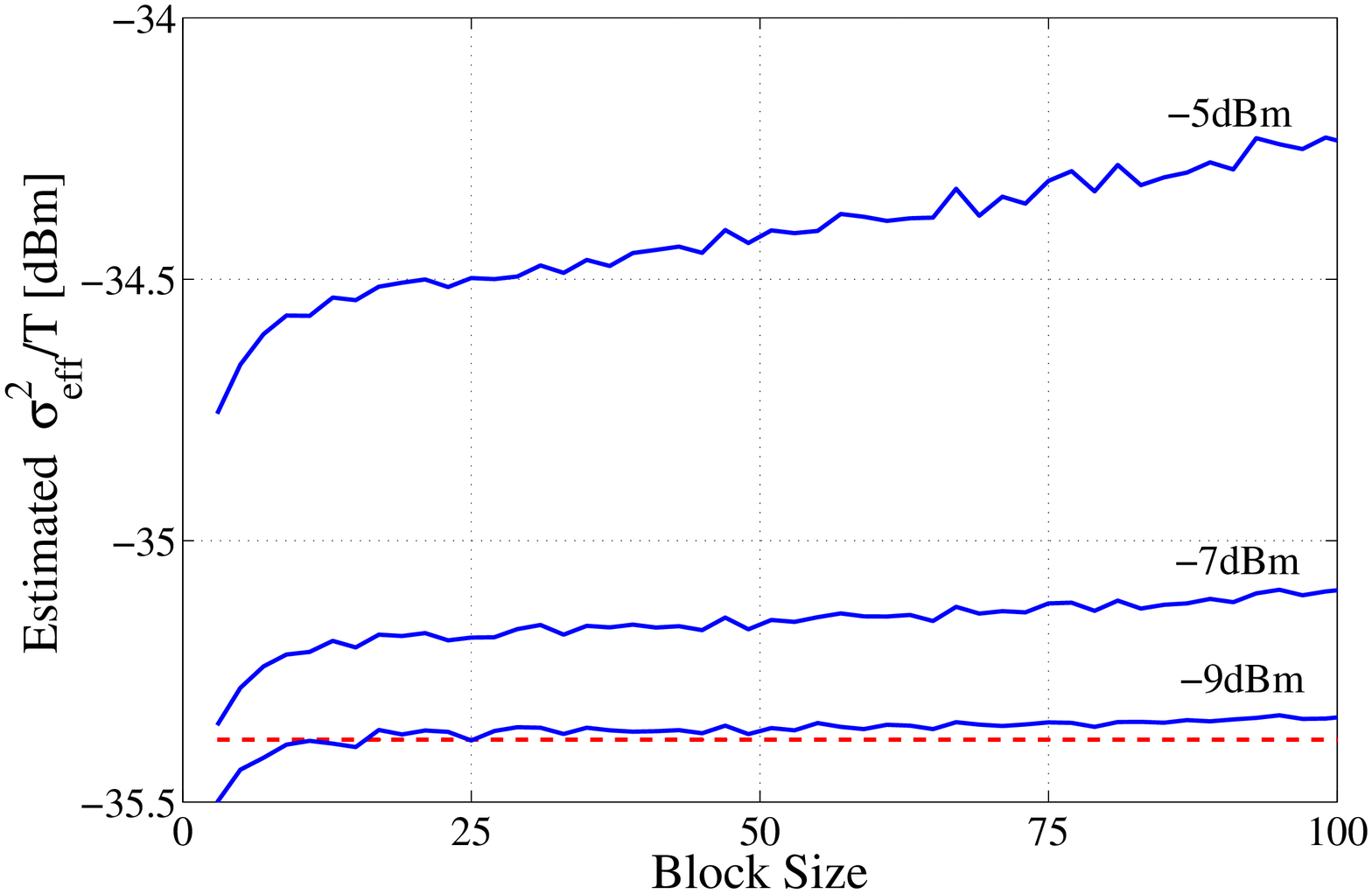,height=0.3\textwidth,width=0.4\textwidth,clip=}
\caption{The numerically estimated $\sigma_{\mathrm{eff}}^2$ (normalized by $T$) vs. block-size in a 500 km link for input average power levels of -9dBm, -7dBm and-5dBm. Red dashed line shows $\sigma_{\mathrm{ASE}}^2/T$. Due to insufficient statistics for small values of $N$, the estimated $\sigma_{\mathrm{eff}}^2$ grows rapidly with block-size. Then, when the accumulated statistics is sufficient, the growth is much slower and it is due to the fact that phase fluctuations inflate the estimated $\sigma_{\mathrm{eff}}^2$.}
\label{fig2eqwfewfew3}
\end{figure}

\begin{figure*}[t]
\center
\hspace{0.5cm}
\begin{subfigure}[t]{0.4\textwidth}
\includegraphics[height=0.76\textwidth,width=1\textwidth]{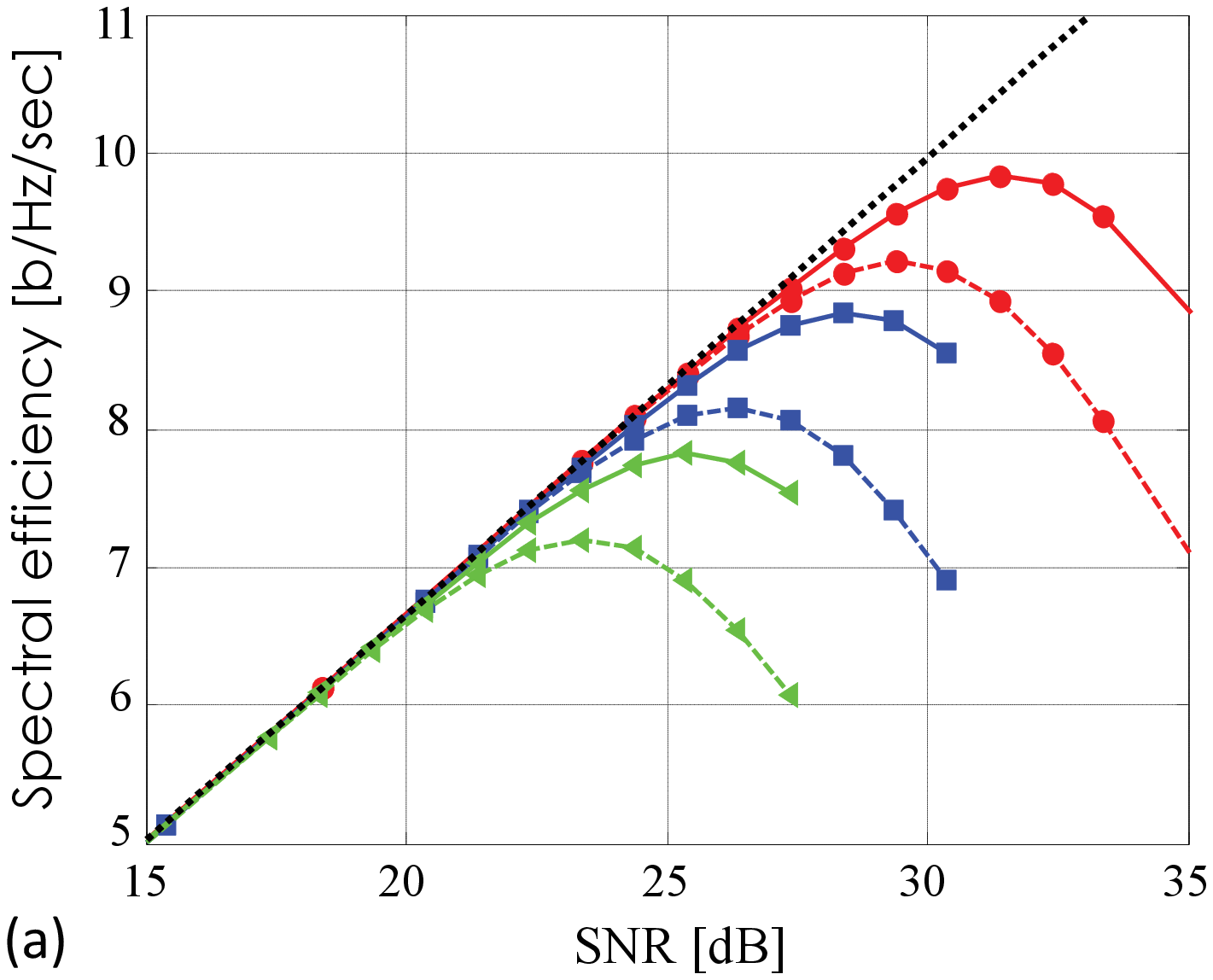}
\label{fewj90}
\end{subfigure}\hspace{0.5cm}
\begin{subfigure}[t]{0.4\textwidth}
\includegraphics[height=0.76\textwidth,width=1.06\textwidth]{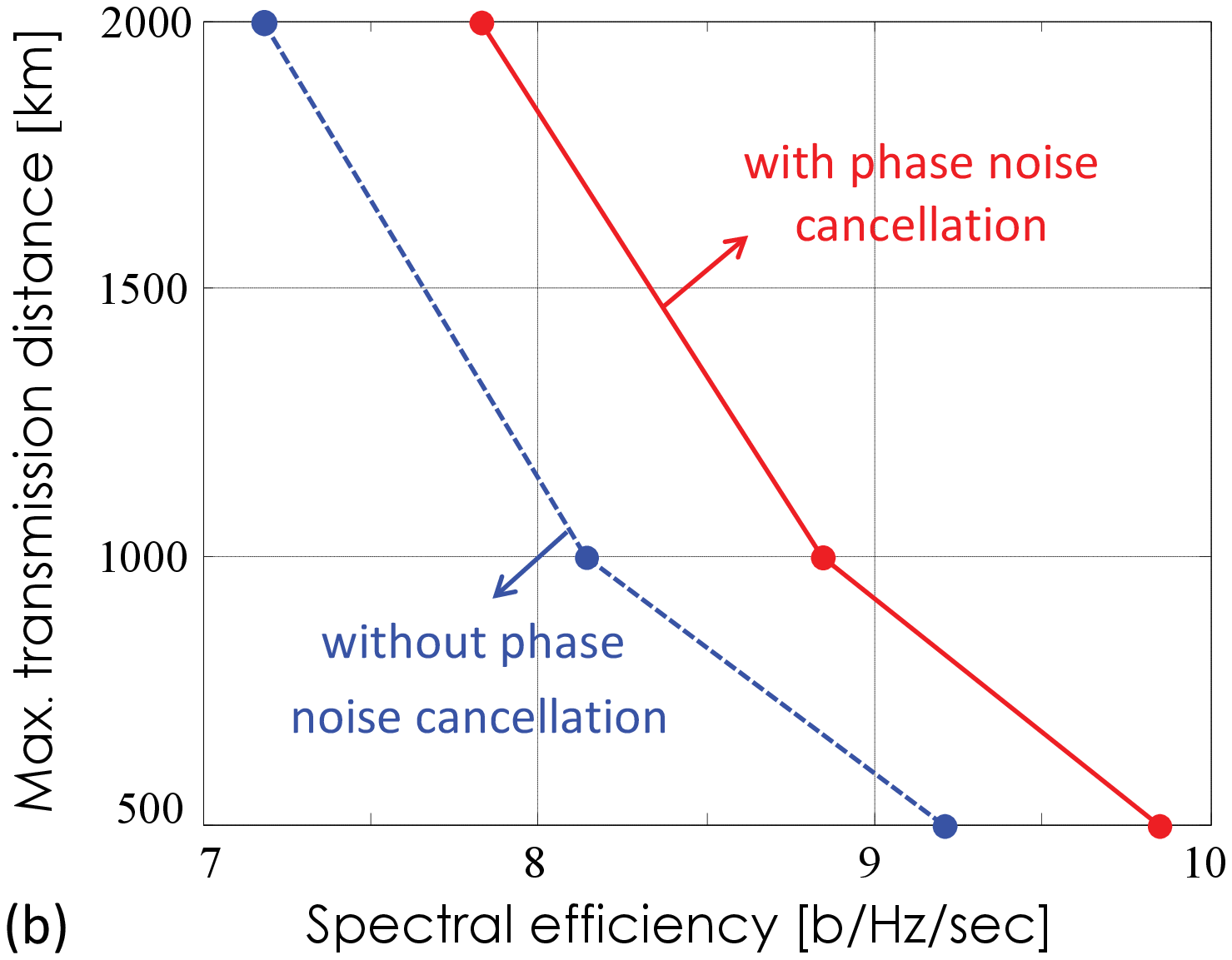}
\label{2d32e2d}
\end{subfigure}

\vspace{-0.5cm}
\center
\begin{subfigure}[t]{0.4\textwidth}
\includegraphics[height=0.76\textwidth,width=1\textwidth]{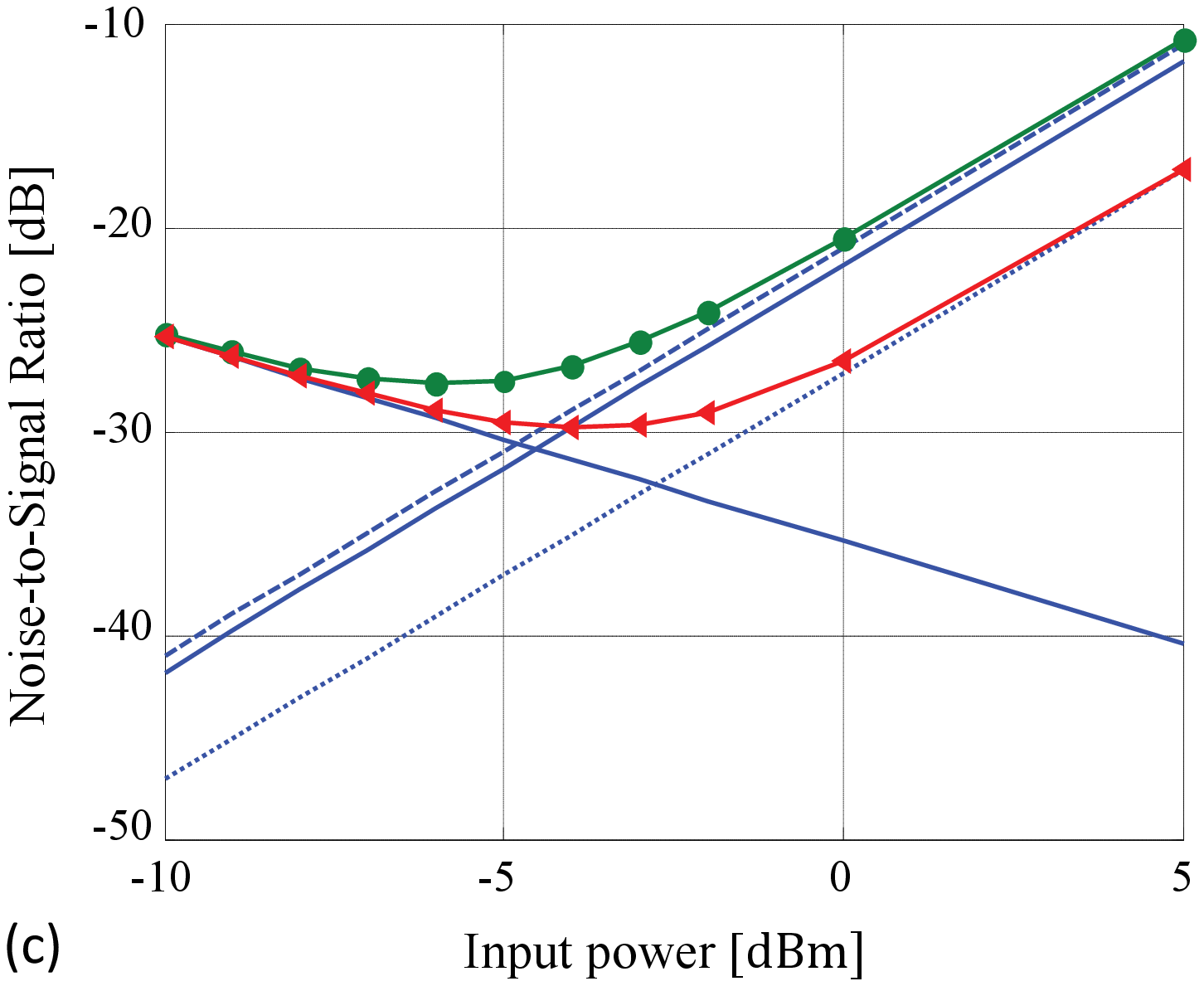}
\label{csa421}
\end{subfigure}\hspace{0.5cm}
\vspace{-0.5cm}
\caption{(a) Capacity lower bound vs. linear SNR for $500$km (red dots), $1000$km (blue squares), and $2000$km (green triangles). Dashed curves result from treating the entire nonlinear noise as noise and they coincide with \cite{CapacityLimitsofOpticalFiberNetworks}. Solid curves represent the new bounds derived here. Dotted curve represents the Shannon limit $\log_2 (1+\text{SNR})$. (b) The maximum achievable transmission distance as a function of spectral efficiency with (solid) and without (dashed) phase-noise separation. (c) Noise to signal ratio vs. average input power in a 500 km link. Decreasing solid line shows $\sigma_{\mathrm{ASE}}^2/(PT)$, increasing solid line shows $\sigma_{\mathrm{\theta}}^2$. Dashed line is the theoretical expression for $\sigma_{\mathrm{\theta}}^2$ found in \cite{PseudolinearCapacityMecozzi}. Dotted line is ${\sigma}^2_{\mathrm{NL}}/(PT)$. Triangles and dots show $\sigma^2_{\mathrm{eff}}/(PT)$ with and without phase-noise cancelation, respectively.}\label{ddmo43}

\end{figure*}

In order to extract the angle $\theta_j$ we exploit the fact that the nonlinear phase noise changes very slowly on the scale of the symbol duration. With this being the case we evaluated $\exp(i\theta_j)$ by averaging the variable $x_j^*y_j$ over $N=50$ adjacent symbols and then normalizing the absolute value of the averaged quantity to 1, so as to ensure that we are only extracting phase noise. The estimate of $\exp(i\theta_j)$ will be denoted by $\exp(i\hat\theta_j)$ in what follows. We then subtracted $x_j\exp(i\hat\theta_j)$ from $y_j$ to obtain $n_j^{\mathrm{NL}}+n_j$ and to evaluate $\sigma_{\mathrm{eff}}^2$. Note that the choice of $N$ affects the estimated noise variance in two ways. On the one hand, the estimation of $\sigma_{\mathrm{eff}}^2$ improves as $N$ increases (the mean square error is proportional to $N^{-1}$). On the other hand, the assumption of constant phase noise becomes less accurate as $N$ increases. As a result the variations of $\theta_j$ inflate the estimate of $\sigma_{\mathrm{eff}}^2$ and reduce the tightness of our capacity lower bound. Fig. \ref{fig2eqwfewfew3} shows the dependence of the estimated value of $\sigma_{\mathrm{eff}}^2$ on the assumed block-size $N$ for several values of average signal power per-channel. The various curves share an important and very distinct feature. In all cases, the estimated value of $\sigma_{\mathrm{eff}}^2$ grows with $N$ at small $N$ values and then it abruptly transitions to a much slower rate of growth. The fast growth in the first stage is due to the lack of sufficient statistics at small $N$ values, whereas the slow growth in the second stage is due to the slow variations in $\theta_j$ whose significance increases with increasing block-size. Our choice of $N=50$ is always higher than the value of $N$ that corresponds to the transition between the two growth rates, thereby guaranteeing that sufficient statistics is used in all cases (albeit at the expense of a slightly overestimated $\sigma_{\mathrm{eff}}^2$).
Finally, the variance $\sigma_\theta^2$ was evaluated by extracting $\hat\theta_j$ from a sliding window average (of width $N=50$) performed over all simulated symbols. This procedure was used in order to numerically estimate the ACF of $\theta_j$, and the result which is plotted by the solid red curve in Fig. \ref{fig2eqw23}, is in good agreement with the corresponding analytical curve.

Fig. \ref{ddmo43}a shows the capacity lower bound curves as a function of the linear SNR (which is the ratio between the average signal power and the power of the ASE noise within the channel bandwidth). The dashed curves correspond to the case in which we do not separate the phase noise and treat the entire nonlinear distortions as noise. These curves accurately reproduce the results of \cite{CapacityLimitsofOpticalFiberNetworks}. The solid curves represent our new lower bound, achieving a peak capacity that is higher by approximately 0.7 bit/sec/Hz in all cases. In Fig. \ref{ddmo43}b we invert the peak capacity results of Fig. \ref{ddmo43}a so as to plot the maximum achievable system length as a function of the spectral efficiency. As is evident from the figure, the achievable system length is approximately doubled by exploiting our scheme. Finally in Fig. \ref{ddmo43}c we show the various noise contributions as a function of the average power per channel in the case of a 500 km link. The monotonically decreasing blue curve shows the noise to signal ratio (NSR) $\sigma_{\mathrm{ASE}}^2/(PT)$ due to the ASE noise by itself. The monotonically increasing solid blue curve shows $\sigma_\theta^2$, which in the limit of small  variations in $\theta_j$, represents the NSR due to phase noise. The dashed blue line is the theoretical expression for $\sigma_\theta^2$ as given in \cite{PseudolinearCapacityMecozzi} (see Eq. \eqref{eqcsdcf834}), which is seen to be in very good agreement with our numerical result. The blue dotted line shows the NSR due to the residual nonlinear noise $\sigma_{\mathrm{NL}}^2/(PT)$ after separating the phase noise. As is evident in the figure, separation of nonlinear phase-noise reduces the nonlinear noise by approximately 6dB. Triangles and dots show $\sigma^2_{\mathrm{eff}}/(PT)$ with and without phase-noise cancelation, respectively. Evidently, the minimum effective NSR of the system is improved by approximately 2dB. We note that to facilitate the distinction between the noise contributions the simulation that produced Fig. \ref{ddmo43}c was performed without ASE propagation. In Figs. \ref{ddmo43}a and \ref{ddmo43}b ASE noise was propagated, although similar results were observed when the ASE was added at the end.

\vspace{.4pc}\noindent{\bf 4. Conclusions}\vspace{.4pc}\newline
We derived a new lower bound for the capacity of the nonlinear fiber channel. By taking into account the fact that phase-noise is one of the most significant consequences of nonlinear interference, and by taking advantage of the fact that this noise is characterized by strong temporal correlations we showed that one can increase the peak capacity per polarization by approximately 0.7bit/s/Hz. Equivalently, we showed that the length of a system can be (almost) doubled for a given transmission rate.

\bibliographystyle{IEEEtran}
\bibliography{IEEEabrv,mybib2}

%
%
%
%
%
%
%

\end{document}